\documentclass[11pt,leqno,fleqn]{article}
\newenvironment{ISItext}
{\normalsize\rm\setlength{\parindent}{1cm}\setlength{\parskip}{0pt}}
{\vskip 12pt}

\begin{document}
\newcommand{\ISItitle}[1]{\vskip 0pt\setlength{\parindent}{0cm}\Large\textbf{#1}\vskip 12pt}
\newcommand{\ISIsubtitleA}[1]{\normalsize\rm\setlength{\parindent}{0cm}\textbf{#1}\vskip 12pt}
\newcommand{\ISIsubtitleB}[1]{\normalsize\rm\setlength{\parindent}{0cm}\textbf{#1}\vskip 12pt}
\newcommand{\ISIsubtitleFig}[1]{\normalsize\rm\setlength{\parindent}{0cm}
\textbf{\textit{#1}}\vskip 12pt}
\newcommand{\ISIauthname}[1]{\normalsize\rm\setlength{\parindent}{0cm}#1 \\}
\newcommand{\ISIauthaddr}[1]{\normalsize\rm\setlength{\parindent}{0cm}\it #1 \vskip 12pt}

\ISItitle{Detection of two-sided alternatives in a Brownian Motion
model\footnote{This research was supported in part by the U.S. National Science
Foundation under Grant No. CCR-02-05214.}}

\ISIauthname{Hadjiliadis, Olympia (1st author)}
\ISIauthaddr{Princeton University, Department of Electrical Engineering\\
Engineering Quadrangle, Olden Street\\
Princeton, NJ 08544 U.S.A.\\
E-mail: ohadjili@princeton.edu}

\ISIauthname{Poor, H. Vincent (2nd author)}
\ISIauthaddr{Princeton University, Department of Electrical Engineering\\
Engineering Quadrangle, Olden Street\\
Princeton, NJ 08544 U.S.A.\\
E-mail: poor@princeton.edu}

\ISIsubtitleA{Introduction and motivation}

\begin{ISItext}
The need for statistical surveillance has been noted in many different areas,
including quality control (see for example \cite{Basseville:1993}), epidemiology
(see for example \cite{Weatherall:1976}), medicine (see for example
\cite{Frisen:1992}), machinery monitoring, seismology, finance (see for example
\cite{Andersson:2002}) etc. In this work, we address the problem of the detection of
two-sided alternatives in a Brownian motion model. This model is the continuous time
equivalent to the discrete time Gaussian observation model. For stochastic systems
with linear dynamics and linear observations that are driven by Gaussian noise, the
Kalman-Bucy innovation process is known to be a sequence of independent Gaussian
random variables. Such models can be used to study systems subject to system
component failures and other systems involving small non-linearities
(\cite{WJ:1976,THS:1977}). Fault detection in a navigation system, where an abrupt
change in the model parameters corresponds to an abrupt change in the mean of the
Kalman filter innovations is an instance of such a situation
(\cite{Tartakovsky:1993,Basseville:1993}). The sign of the change depends on the
signs of the gyro errors (\cite{WDC:1975}). Another instance of such a model can be
seen in sensor failure detection for the monitoring of traffic incidents on
freeways. Each sensor is placed in different locations on the freeway and records
the mean velocity and density of cars. An abrupt and systematic change in these
recordings would trigger an abrupt change in the Kalman filter innovations in either
direction depending on whether the sensor is consistently overestimating or
underestimating (see \cite{WCGGHK:1980}). Identification and removal of the faulty
sensor becomes essential. The continuous version of the Kalman filter innovations in
all of the above linear Gaussian models is seen to be a Brownian motion
(\cite{Oksendal:1985}). Other applications includes the detection of a rhythm jump
of the heartbeat during an ECG (see \cite{DTW:1986}) and in the detection of a
positive or negative drift in the log of stock price dynamics.

This paper is concerned with the quickest detection of two-sided alternatives in the
drift of a Brownian motion. In particular, we find the best 2-CUSUM stopping rule
with respect to an extended Lorden criterion. Although, the mathematical formulation
is done in the context of the one-dimensional case, extension to the vector case
that corresponds to the Kalman innovations in linear systems described above is
straightforward (see \cite{Moustakides:2004}).

%We begin with the mathematical formulation of the problem. We then proceed to define
%the general 2-CUSUM stopping rule, and to further specify the class of 2-CUSUM
%harmonic mean rules. We  subsequently provide three expressions for the first moment
%of a general 2-CUSUM stopping rule and an expression for the rate of change of this
%first moment with respect to one of the threshold parameters. We proceed to
%establish explicit upper and lower bound on the expected value of a general 2-CUSUM
%stopping rule, and use it to enhance the existing upper and lower bounds that appear
%in the literature (see \cite{Dragalin:1997}). We also establish explicit upper and
%lower bounds on the rate of change of this first moment with respect to one of the
%threshold parameters. In Section 6, we prove the uniqueness of the best 2-CUSUM
%stopping rule, both in the case of a symmetric change and in the case of a
%non-symmetric change. Finally, in Section 7, we conclude with some closing remarks.

\pagebreak

\ISIsubtitleB{Mathematical formulation and main results}

We sequentially observe a process $\{\xi_t\}$ with the following dynamics:
\begin{displaymath}
\label{theproblem} d\xi_t = \left \{
\begin{array}{ll}
\hspace*{1ex} dw_t  & t \le \tau \\
[6ex]

\begin{array}{l}
\mu_1 dt + dw_t \\
\hspace*{5ex}\textrm{{\small or}} \\
\hspace*{-2ex}-\mu_2 dt + dw_t
\end{array}

&  t \ge \tau

\end{array}
\right.
\end{displaymath}
where $\tau$, the time of change, is assumed to be deterministic but unknown; $w_t$
is a standard Brownian motion process; $\mu_i$, the possible drifts to which the
process can change, are assumed to be known, but the specific drift to which the
process is changing is unknown. Both $\mu_1$ and $\mu_2$ are assumed to be positive.

The probability triplet consists of $(C[0,\infty],\cup_{t>0}{\cal F}_t)$, where
${\cal F}_t=\sigma\{\xi_s, 0<s \le t\}$ and the families of probability measures
$\{\mathcal{P}_\tau^i\},~\tau \in [0,\infty)$, whenever the change is
$\mu_i,~i=1,2,$ and $\mathcal{P}_\infty$, the Wiener measure.

Our goal is to detect a change by means of a stopping rule $T$ adapted to the
filtration $\mathcal{F}_t$. As a performance measure for this stopping rule we
propose an extended Lorden  criterion (see \cite{Hadjil:2006})
\begin{equation}
J_L(T)=\max_i \sup_{\tau}~ \textrm{essup}~ E_{\tau}^i \left[ {(T-\tau)}^{+}
|{\cal{F}}_\tau \right]. \label{JL}
\end{equation}
This gives rise to the following min-max constrained optimization problem:
\begin{eqnarray}
\inf_T J_L(T) & \ & \nonumber
\\
\textrm{subject to }E_\infty \left[T \right] \geq \gamma, & \ & \label{eqnproblem}
\end{eqnarray}

where the constraint specifies the minimum allowable mean time between false alarms.

%As discussed first by Moustakides (\cite{Moustakides:1986}), and later by Poor
%(\cite{Poor:1998}), in seeking solutions to the above problem, we can restrict our
%attention to stopping times that achieve the false alarm constraint with equality.
%This follows from the fact that, if $E_\infty\left[T\right]>\gamma$, we can produce
%a stopping time that achieves the constraint with equality without increasing the
%detection delay, simply by randomizing between $T$ and the stopping time that is
%identically $0$. We, therefore, look for stopping rules that are
%${\cal{F}}_t$-adapted $\forall~t$, and that satisfy the false alarm constraint with
%equality in order to find a solution to (\ref{eqnproblem}).
%

In this paper we seek the best 2-CUSUM stopping rule in the sense described in
(\ref{eqnproblem}). The 2-CUSUM rules have been proposed and used extensively due
not only to the simplicity in the calculation of their first moment(see
\cite{Siegmund:1985}), but also to their asymptotically optimal character (see
\cite{Hadjil:2006}, \cite{Tartakovsky:1994}).

We begin by defining the CUSUM statistics and stopping rules of interest.

\noindent{\bf Definition} Let $\nu_1>0$ and $\nu_2>0$. Define
\begin{enumerate}
\item
$u_t^+=\frac{\log\frac{dP_0^1}{dP_\infty}|\mathcal{F}_t}{\mu_1}=\xi_t-\frac{1}{2}\mu_1
t$; $m_t^+=\inf_{s\le t}u_s^+$; $y_t^+=u_t^+-m_t^+$,
\item
$u_t^-=\frac{\log\frac{dP_0^2}{dP_\infty}|\mathcal{F}_t}{\mu_2}=-\xi_t-\frac{1}{2}\mu_2t$;
$m_t^-=\inf_{s \le t}u_s^-$; $y_t^-=u_t^--m_t^-$,
\item $T_1(\nu_1)=\inf\{t>0;y_t^+ \ge \nu_1\}$, and
\item $T_2(\nu_2)=\inf\{t>0;y_t^- \ge \nu_2\}$.
\end{enumerate}

The 2-CUSUM stopping rules are then of the form $T(\nu_1,\nu_2)=T_1(\nu_1) \wedge
T_2(\nu_2)$.

We also define the following stopping rules, the use of which will become apparent
later.

\noindent{\bf Definition} \label{U} For $a>0$ and $b>0$, we define
\begin{enumerate}
\item $U^+(a)=\inf\{t>0; u_t^+\ge a\}$,
\item $U^-(b)=\inf\{t>0; -u_t^-\le -b\}$, and
\item $\Pi(a,b)=P\left(U^+(a)<U^-(b)\right).$
\end{enumerate}
%For reasons that will become apparent, we introduce the following definition.
%
%\noindent{\bf Definition} \label{additional} Define
%\begin{enumerate}
%\item $u_{t-}^-=u_t^-+\frac{1}{2}(\mu_1+\mu_2)t$,
%\item $m_{t-}^-=\inf_{s \le t} u_{s_-}^-$,
%\item $y_{t-}^-=u_{t-}-m_{t-}$ and
%\item $T_2^-(\nu_2)=\inf\{t>0;y_{t-}^-\ge \nu_2\}$.
%\end{enumerate}

%At this point, let us remind ourselves of the delay function, as it appears in
%\cite{Hadjil:2006} and in \cite{Siegmund:1985}. We have
%\begin{eqnarray}
%\label{ItoINFTYT1} E_\infty(T_1(\nu_1)) & = & 2f_{\nu_1}(\mu_1),
%\\
%\label{ItoINFTYT2} E_\infty(T_2(\nu_2)) & = & 2f_{\nu_2}(\mu_2),
%\\
%\label{Ito1T1}E_0^1(T_1(\nu_1)) & = & 2f_{\nu_1}(-\mu_1),
%\\
%\label{Ito1T2} E_0^1(T_2(\nu_2)) & = & 2f_{\nu_2}(\mu_2+2\mu_1),
%\\
%\label{Ito2T1} E_0^2(T_1(\nu_1)) & = & 2f_{\nu_1}(\mu_1+2\mu_2),
%\end{eqnarray}
%and \begin{eqnarray} \label{Ito2T2} E_0^2(T_2(\nu_2)) & = & 2f_{\nu_2}(-\mu_2),
%\end{eqnarray}
%where $f_x(y)=\frac{e^{yx}-yx-1}{y^2}$.

For any 2-CUSUM stopping rule $T$ we have (see \cite{Hadjil:2006}) $J_L(T) =
\max\{E_0^1\left[T\right],E_0^2\left[T\right]\}$.

We now classify 2-CUSUM rules according to the class
$\mathcal{G}=\{T(\nu_1,\nu_2);\nu_1=\nu_2\}$ of harmonic mean rules and the classes
$\mathcal{C}_1= \{T(\nu_1,\nu_2)~|~\nu_1>\nu_2>0\}$ and $\mathcal{C}_2=
\{T(\nu_1,\nu_2)~|~\nu_2>\nu_1>0\}$ of non-harmonic mean rules. For simplicity of
display and notation we finally define the constants $m=\min\{\nu_1,\nu_2\}$,
$M=\max\{\nu_1,\nu_2\}$ and the functions $C_m(x,y)=\frac{f_m(x)^2}{f_m(x)+f_m(y)}$,
$\lambda_x(y)=\frac{1}{yf_x(y)+x}$, $f_y^*(x)=f_x(y)=\frac{e^{yx}-yx-1}{y^2}$. We
now summarize the main results.
%Finally, for $T \in \mathcal{C}_i$ for $i=1,2$, let $R_1^i=\frac{d
%E_0^1\left[T\right]}{dM}$ and $R_2^i=\frac{d E_0^2\left[T\right]}{dM}$.

\vspace{1ex}

\noindent {\bf Theorem} \label{nonharmonicmean} Let $T(\nu_1,\nu_2)=T_1(\nu_1)
\wedge T_2(\nu_2)$ be any 2-CUSUM stopping rule and denote $T(\nu_1,\nu_2)$ by $T$.
Then, the following is true under any of the measures $P_\infty$, $P_0^1$ and
$P_0^2$:
\begin{enumerate}
\item for all $T \in \mathcal{C}_1$, $m=\nu_2$, $M=\nu_1$, we have
\begin{eqnarray*}
\label{expvaluev2lessv1} \nonumber E\left[T\right]& = &
E\left[T_2(m)\right]\cdot\left[1-\frac{E\left[T_2(m)\right]}{E\left[T_1(m)\right]+E\left[T_2(m)\right]}\lim_{n
\to \infty}\Pi\left(\frac{1}{n},m\right)^{(M-m)n} \right],
\end{eqnarray*}
and
\item for all $T \in \mathcal{C}_2$, $m=\nu_1$, $M=\nu_2$,  we have
\begin{eqnarray*}
\label{expvaluev1lessv2} \nonumber E\left[T\right] & =
&E\left[T_1(m)\right]\cdot\left[1-\frac{E\left[T_1(m)\right]}{E\left[T_1(m)\right]+E\left[T_2(m)\right]}\lim_{n
\to \infty}\left(1-\Pi(m,\frac{1}{n})\right)^{(M-m)n}\right].\end{eqnarray*}
\end{enumerate}

\noindent{\bf Corollary} \label{ULbound} Let $T(\nu_1,\nu_2)=T_1(\nu_1) \wedge
T_2(\nu_2)$ be any 2-CUSUM stopping rule and denote $T(\nu_1,\nu_2)$ by $T$. Then,
for all $T \in \mathcal{C}_1$, $m=\nu_2$, $M=\nu_1$ and
\begin{eqnarray}
\label{UBv1v2INFTY} E_\infty\left[T\right]& \le &
2f_{m}(\mu_2)\cdot\left[1-\frac{C_m(\mu_2,\mu_1)}{f_m(\mu_2)} e^{-\lambda_{m}(-\mu_1)(M-m)} \right], \\
\label{LBv1v2INFTY} E_\infty\left[T\right]& \ge &
2f_{m}(\mu_2)\cdot\left[1-\frac{C_m(\mu_2,\mu_1)}{f_m(\mu_2)} e^{-\lambda_{m}(\mu_2)(M-m)} \right], \\
\label{UBv1v201} E_0^1\left[T\right]& \le &
2f_{m}(\mu_2+2\mu_1)\cdot\left[1-\frac{C_m(\mu_2+2\mu_1,-\mu_1)}{f_m(\mu_2+2\mu_1)}(e^{-\lambda_{m}(\mu_1)(M-m)} \right], \\
\label{LBv1v201} E_0^1\left[T\right]& \ge &
2f_{m}(\mu_2+2\mu_1)\cdot\left[1-\frac{C_m(\mu_2+2\mu_1,-\mu_1)}{f_m(\mu_2+2\mu_1)} e^{-\lambda_{m}(\mu_2+2\mu_1)(M-m)} \right], \\
\label{UBv1v202} E_0^2\left[T\right]& \le &
2f_{m}(-\mu_2)\cdot\left[1-\frac{C_m(-\mu_2,\mu_1+2\mu_2)}{f_m(-\mu_2)}e^{-\lambda_{m}\left(-(\mu_1+2\mu_2)\right)(M-m)} \right], \textrm{ and }\\
\label{LBv1v202} E_0^2\left[T\right]& \ge &
2f_{m}(-\mu_2)\cdot\left[1-\frac{C_m(-\mu_2,\mu_1+2\mu_2)}{f_m(-\mu_2)}e^{-\lambda_{m}(-\mu_2)(M-m)}
\right].
\end{eqnarray}
%and
%\item for all $T \in \mathcal{C}_2$, $m=\nu_1$, $M=\nu_2$ and
%\begin{eqnarray}
%\label{UBv2v1INFTY} E_\infty\left[T\right]& \le &
%2f_{m}(\mu_1)\cdot\left[1-\frac{C_m(\mu_1,\mu_2)}{f_m(\mu_1)} e^{-\lambda_{m}(-\mu_2)(M-m)} \right], \\
%\label{LBv2v1INFTY} E_\infty\left[T\right]& \ge &
%2f_{m}(\mu_1)\cdot\left[1-\frac{C_m(\mu_1,\mu_2)}{f_m(\mu_1)} e^{-\lambda_{m}(\mu_1)(M-m)} \right], \\
%\label{UBv2v101} E_0^1\left[T\right]& \le &
%2f_{m}(-\mu_1)\cdot\left[1-\frac{C_m(-\mu_1,\mu_2+2\mu_1)}{f_m(-\mu_1)}e^{-\lambda_{m}\left(-(2\mu_1+\mu_2)\right)(M-m)} \right], \\
%\label{LBv2v101} E_0^1\left[T\right]& \ge &
%2f_{m}(-\mu_1)\cdot\left[1-\frac{C_m(-\mu_1,\mu_2+2\mu_1)}{f_m(-\mu_1)}e^{-\lambda_{m}(-\mu_1)(M-m)} \right], \\
%\label{UBv2v102} E_0^2\left[T\right]& \le &
%2f_{m}(\mu_1+2\mu_2)\cdot\left[1-\frac{C_m(\mu_1+2\mu_2,-\mu_2)}{f_m(\mu_1+2\mu_2)}e^{-\lambda_{m}(\mu_2)(M-m)} \right], \\
%\label{LBv2v102} E_0^2\left[T\right]& \ge &
%2f_{m}(\mu_1+2\mu_2)\cdot\left[1-\frac{C_m(\mu_1+2\mu_2,-\mu_2)}{f_m(\mu_1+2\mu_2)}e^{-\lambda_{m}(\mu_1+2\mu_2)(M-m)}
%\right].
%\end{eqnarray}
%\end{enumerate}
Similar results hold for $T \in \mathcal{C}_2$. For more details please refer to
\cite{HadjiliadisPoor:2007}.

\noindent{\bf Theorem} \label{main} The best $T^*$ 2-CUSUM stopping rule exists and
is unique and we distinguish the following cases
\begin{enumerate}
\item If $\mu_1<\mu_2$ then $T^*\in \mathcal{C}_2$.
\item If $\mu_2<\mu_1$ then $T^*\in \mathcal{C}_1$.
\item If $\mu_1=\mu_2$ then $T^*\in \mathcal{G}$.
\end{enumerate}

We refer the reader to \cite{HadjiliadisPoor:2007} for a detailed proof of all of
the above results and other interesting corollaries.

%\noindent{\bf Corollary} \label{expr} Let $T(\nu_1,\nu_2)=T_1(\nu_1) \wedge
%T_2(\nu_2)$ be any 2-CUSUM stopping rule and denote $T(\nu_1,\nu_2)$ by $T$. Then,
%\begin{enumerate}
%\item for all $T \in \mathcal{C}_1$, $m=\nu_2$, $M=\nu_1$ and
%\begin{eqnarray}
%\label{EQv1v2INFTY} E_\infty\left[T\right]& = &
%2f_{m}(\mu_2)\cdot\left[1-\frac{C_m(\mu_2,\mu_1)}{f_m(\mu_2)} e^{-\lambda_{m}(x_\infty)(M-m)} \right], \\
%\label{EQv1v201} E_0^1\left[T\right]& = &
%2f_{m}(\mu_2+2\mu_1)\cdot\left[1-\frac{C_m(\mu_2+2\mu_1,-\mu_1)}{f_m(\mu_2+2\mu_1)}e^{-\lambda_{m}(x_1)(M-m)} \right], \\
%\label{EQv1v202} E_0^2\left[T\right]& = &
%2f_{m}(-\mu_2)\cdot\left[1-\frac{C_m(-\mu_2,\mu_1+2\mu_2)}{f_m(-\mu_2)}
%e^{-\lambda_{m}(x_2)(M-m)} \right],
%\end{eqnarray}
%where
%\begin{eqnarray*}
%-\mu_1<& x_\infty< & \mu_2, \\
%\mu_1 < & x_1< &\mu_2+2\mu_1, \\
%-(\mu_1+2\mu_2)< & x_2< &-\mu_2,
%\end{eqnarray*}
%and
%\item for all $T \in \mathcal{C}_2$, $m=\nu_1$, $M=\nu_2$ and
%\begin{eqnarray}
%\label{EQv2v1INFTY} E_\infty\left[T\right]& = &
%2f_{m}(\mu_1)\cdot\left[1-\frac{C_m(\mu_1,\mu_2)}{f_m(\mu_1)} e^{-\lambda_{m}(y_\infty)(M-m)} \right], \\
%\label{EQv2v101} E_0^1\left[T\right]& = &
%2f_{m}(-\mu_1)\cdot\left[1-\frac{C_m(-\mu_1,\mu_2+2\mu_1)}{f_m(-\mu_1)}e^{-\lambda_{m}(y_1)(M-m)} \right], \\
%\label{EQv2v102} E_0^2\left[T\right]& = &
%2f_{m}(\mu_1+2\mu_2)\cdot\left[1-\frac{C_m(\mu_1+2\mu_2,-\mu_2)}{f_m(\mu_1+2\mu_2)}e^{-\lambda_{m}(y_2)(M-m)}
%\right],
%\end{eqnarray}
%where
%\begin{eqnarray*}
%-\mu_2 < & y_\infty < & \mu_1, \\
%-(2\mu_1+\mu_2)< & y_1< & -\mu_1, \\
%\mu_2< & y_2< & \mu_1+2\mu_2.
%\end{eqnarray*}
%\end{enumerate}

\end{ISItext}

%\ISIsubtitleFig{Figure or Table Title}
%
%\begin{ISItext}
%Body Text Style 1. Body Text Style 1. Body Text Style 1. Body Text Style 1.
%\end{ISItext}
%
%

%\ISIsubtitleB{REFERENCES (R\'EFERENCES)}

\bibliographystyle{plain}
\bibliography{references}

\begin{thebibliography}{10}

\bibitem{Andersson:2002}
E.~Andersson.
\newblock Monitoring cyclical processes: {A} non-parametric approach.
\newblock {\em Journal of Applied Statistics}, 29:973--990, 2002.

\bibitem{Basseville:1993}
M.~Basseville and I.~Nikiforov.
\newblock {\em Detection of Abrupt Changes: Theory and Applications.}
\newblock Prentice Hall, Englewood Cliffs, NJ, 1993.

\bibitem{DTW:1986}
P.C. Doerschuk, R.~R. Tenney, and A.~S. Willsky.
\newblock {\em Estimation-based {A}pproaches to {R}hythm {A}nalysis in
  Electrocardiograms}, {L}ecture {N}otes in {C}ontrol and {I}nformation
  {S}cience III, pages 297--313.
\newblock Springer-Verlag, New York, 1986.

\bibitem{Frisen:1992}
M.~Frisen.
\newblock Evaluations of methods for statistical surveillance.
\newblock {\em Statistics in Medicine}, 11:1489--1502, 1992.

\bibitem{Hadjil:2006}
O.~Hadjiliadis and G.~V. Moustakides.
\newblock Optimal and asymptotically optimal {CUSUM} rules for change point
  detection in the {B}rownian motion model with multiple alternatives.
\newblock {\em Theory of {P}robability and {I}ts {A}pplications (Teoriya
  Veroyatnostei i ee Primeneniya)}, 50(1):131--144, 2006.

\bibitem{HadjiliadisPoor:2007}
O.~Hadjiliadis and H.V. Poor.
\newblock The best 2-{CUSUM} stopping rules for quickest detection of two-sided
  alternatives in a {B}rownian motion model.
\newblock 2007.
\newblock submitted.

\bibitem{Moustakides:2004}
G.~V. Moustakides.
\newblock Optimality of the {CUSUM} procedure in continuous time.
\newblock {\em Annals of Statistics}, 32(1):302--315, 2004.

\bibitem{Oksendal:1985}
B.~Oksendal.
\newblock {\em Stochastic differential equations}.
\newblock Springer-Verlag, Berlin, 1985.

\bibitem{Siegmund:1985}
D.~Siegmund.
\newblock {\em Sequential Analysis}.
\newblock Springer-Verlag, New York, 1st edition, 1985.

\bibitem{Tartakovsky:1993}
A.~Tartakovsky.
\newblock Minimal time detection algorithms and applications to flight systems.
\newblock December 1993.
\newblock Technical report, Flight {S}ystems {R}esearch {C}enter, {U}niversity
  of {C}alifornia, {L}os {A}ngeles.

\bibitem{Tartakovsky:1994}
A.~G. Tartakovsky.
\newblock Asymptotically minimax multi-alternative sequential rule for disorder
  detection.
\newblock {\em Statistics and Control of Random Processes: Proceedings of the
  Steklov Institute of Mathematics}, 202(4):229--236, 1994.
\newblock American Mathematical Society, Providence, RI.

\bibitem{THS:1977}
R.~R. Tenney, R.S. Hebbert, and N.~R. Sandell.
\newblock A tracking filter for maneuvering sources.
\newblock {\em IEEE {T}ransactions on {A}utomatic {C}ontrol}, 22(2):246--251,
  April 1977.

\bibitem{Weatherall:1976}
J.A.C. Weatherall and J.C. Haskey.
\newblock Surveillance of malformations.
\newblock {\em British {M}edical {B}ulletin}, 32:39--44, 1976.

\bibitem{WCGGHK:1980}
A.S. Willsky, E.~Chow, S.~B. Gershwin, C.~S. Greene, P.K. Houpt, and A.~L.
  Kurkjian.
\newblock Dynamic model-based techniques for the detection of incidents on
  freeways.
\newblock {\em IEEE {T}ransactions on {A}utomatic {C}ontrol}, 25(3):347--359,
  1980.

\bibitem{WDC:1975}
A.S. Willsky, J.J. Deyst, and B.S. Crawford.
\newblock Two self-test methods applied to an inertial system problem.
\newblock {\em Journal of Spacecraft}, 12(7):434--437, 1975.

\bibitem{WJ:1976}
A.S. Willsky and H.~Jones.
\newblock A generalized likelihood ratio approach to the detection and
  estimation of jumps in linear systems.
\newblock {\em IEEE {T}ransactions on {A}utomatic {C}ontrol}, 21(1):108--112,
  February 1976.

\end{thebibliography}

%\begin{ISItext}
%\begin{small}
%Body Text Style 2. Body Text Style 2. Body Text Style 2. Body Text Style 2.
%
%Body Text Style 2. Body Text Style 2. Body Text Style 2. Body Text Style 2.
%\end{small}
%\end{ISItext}

\end{document}